# Homogeneous Large-area Quasi-freestanding Monolayer and Bilayer Graphene on SiC

D. Momeni Pakdehi,*,1 K. Pierz,*,1 S. Wundrack,1 J. Aprojanz,2 T. T. N. Nguyen,3 T. Dziomba,1 F. Hohls,1 A. Bakin,4,5 R. Stosch,1 C. Tegenkamp,2,3, F. J. Ahlers,1 and H. W. Schumacher1

1 Physikalisch-Technische Bundesanstalt, Bundesallee 100, 38116 Braunschweig, Germany
2 Institut für Festkörperphysik, Leibniz Universität Hannover, Appelstraße 2, 30167 Hannover, Germany
3 Institut für Physik, Technische Universität Chemnitz, Reichenhainer Straße 70, 09126 Chemnitz, Germany
4 Institut für Halbleitertechnik, Technische Universität Braunschweig, Hans-Sommer Straße 66, 38106 Braunschweig, Germany
5 Laboratory of Emerging Nanometrology (LENA), Technische Universität Braunschweig, Langer Kamp 6a, 38106 Braunschweig, Germany

ABSTRACT. In this study, we first show that the argon flow during epitaxial graphene growth is an important parameter to control the quality of the buffer and the graphene layer. Atomic force microscopy (AFM) and low-energy electron diffraction (LEED) measurements reveal that the decomposition of the SiC substrate strongly depends on the Ar mass flow rate while pressure and temperature are kept constant. Our data are interpreted by a model based on the competition of the SiC decomposition rate, controlled by the Ar flow, with a uniform graphene buffer layer formation under the equilibrium process at the SiC surface. The proper choice of a set of growth parameters allows the growth of defect-free, ultra-smooth and coherent graphene-free buffer layer and bilayer-free monolayer graphene sheets which can be transformed into large-area high-quality quasi-freestanding monolayer and bilayer graphene (QFMLG and QFBLG) by hydrogen intercalation. AFM, scanning tunneling microscopy (STM), Raman spectroscopy and electronic transport measurements underline the excellent homogeneity of the resulting quasi-freestanding layers. Electronic transport measurements in four-point probe configuration reveal a homogeneous low resistance anisotropy on both μm- and mm scales.

## INTRODUCTION

Quasi-freestanding monolayer graphene (QFMLG) can be fabricated by decoupling of an epitaxially grown buffer layer from the underlying SiC substrate, e.g., via hydrogen intercalation. [1–3] The hydrogen intercalation allows the fabrication of p-type monolayer graphene combined with the advantage of the large-scale graphene epitaxial growth directly on semi-insulating SiC substrates with reduced influence on the atop graphene layer. [4] Hence, this approach offers for potential applications a versatile platform as an alternative to epitaxial graphene (EG) with n-type charge carriers in the pristine state.

State-of-the-art QFMLG can be fabricated with high quality proven by low defect-related D-peak intensities in local Raman measurements and high charge carrier mobilities in transport measurements of micrometer-sized Hall bars. [2,5,6] However, it is quite challenging to obtain homogenous QFMLG over mm or cm areas, as can be obtained with EG. [7–10] One reason is the lower temperature used for buffer layer growth (about 1400 °C, ~1 bar) compared to graphene growth (> 1600 °C, ~1 bar) which limits the carbon supply and surface mass transport and thus, the formation of a coherent large-area buffer layer. Such problems are often observed at the SiC step-edge region where the sublimation rate is strongly enhanced. [11] Moreover, hexagonal SiC shows terraces with inequivalent surface energies and decomposition velocities [12,13] which complicates the epitaxial growth concerning thickness control and coverage. Due to these facts, either incomplete buffer layer coverage (at low growth temperatures) [9] or additional graphene-layer formation (at elevated growth temperatures) [14,15] are common consequences at step edges. Practically, such defected buffer layers prevent a reproducible fabrication of large-area homogeneous QFMLG, which is unfavorable regarding electronic device fabrication.

In this study, we first focus on the influence of mass flow rate of argon which is used as an inert atmosphere for the epitaxial buffer layer and graphene growth processes. After that, we show ultra-smooth buffer layer and graphene monolayer fabricated by taking into account the influence of the Ar flow rate as well as other growth determining parameters. The high quality of the produced samples is further demonstrated after hydrogen intercalation.

Inert gas ($N_2$, Ar, etc.) counter pressure was used for many years to improve SiC sublimation growth which prevents unwanted crystal growth before reaching the optimal growth temperature. [16,17] When the Ar counter pressure was introduced to epitaxial graphene growth by SiC sublimation, it led to a groundbreaking improvement of the homogeneity of epitaxial graphene growth. [4,18] Until now, the Ar mass flow was not considered as an important parameter for graphene growth. Our results show, however, that the SiC decomposition rate can be controlled via the Ar flow without varying total pressure and substrate temperature. Atomic force microscopy (AFM) and Raman spectroscopy measurements prove that optimized Ar mass flow conditions lead to the formation of highly homogenous buffer or graphene layers, and after intercalation to high-quality large-area QFMLG and QFBLG, respectively. The STM measurements demonstrate freestanding graphene layers smoothly bridges over the SiC steps on the adjacent terraces. This is further supported by mm-scale Van der Pauw (VdP) as well as µm-scale nano-four-point probe (N4PP) measurements of millimeter-sized samples with high charge carrier mobilities up to 1300 and 3300 (cm$^2$/Vs) for QFMLG and QFBLG at room temperature, respectively, at room temperature. The measured resistance anisotropy values of 15% for QFMLG and 35% for QFBLG are a significant improvement compared to literature. [19]

### SAMPLE PREPARATION

The experiments were performed on the Si-face of the samples (5 x 10 mm$^2$) cut from a semi-insulating 6H-SiC wafer with a nominal miscut of about -0.06° towards [1$\bar{1}$00] (from II-VI Inc.). The substrates were prepared by liquid phase deposition of polymer adsorbates on the surface as described for the polymer assisted sublimation growth (PASG) technique in Refs. [7–9] Epitaxial growth was carried out in a horizontal inductively heated furnace. [20]

Studying the influence of Ar mass flow rate on the buffer layer growth was conducted on three samples $S_0$, $S_{100}$, and $S_{1000}$. After vacuum annealing at 900 °C the buffer layer was grown at 1400 °C (900 mbar Ar atmosphere, 30 min) under Ar mass flow rates of 0, 100 and 1000 sccm, respectively.

The impact of the Ar mass flow rate on the graphene growth was investigated on two samples $G_0$ and $G_{20}$ grown at 1750 °C (Ar atm., 900 mbar, 6 min), under zero and 20 sccm Ar mass flow, respectively. All other parameters were kept constant.

The optimized buffer layer sample $B_0$ was grown under 0 sccm Ar flow (900 mbar) by an annealing procedure with temperature steps at 1200 (10 min), 1400 (5 min) and 1500 °C (5 sec).

Finally, we conducted hydrogen intercalation on $B_0$ (buffer layer) at 900 °C (60 min) to obtain freestanding monolayer graphene (QFMLG). Similarly, quasi-freestanding bilayer graphene (QFBLG) was achieved by hydrogen intercalation on $G_0$ (epi-Graphene) at 1050 °C (2 h). The intercalation was done in hydrogen (5%) and argon (95%) gas mixture (1000 mbar). The optimal temperature was determined by Raman spectroscopy, see supporting information.

### RESULTS AND DISCUSSION

**Influence of Ar flow rate in epitaxial buffer layer/ graphene growth.** The surface morphology of the buffer layer samples grown under different Ar mass flow rates ($S_0$, $S_{100}$ and $S_{1000}$) are plotted in Figure 1. Figure 1a and b show for sample $S_0$ (zero Argon flow) a smooth surface with regular terraces and step heights of 0.75 nm. The clear $(6\sqrt{3} \times 6\sqrt{3})R30°$ spot profile analysis low-energy electron diffraction (SPA-LEED) pattern in Figure 1c indicates the formation of buffer layer graphene which homogenously covers the terraces as shown by the even phase contrast. The homogenous buffer layer growth is attributed to the PASG growth which favors buffer layer nucleation over the entire terrace. [7,9] The different phase contrasts (lighter colors) along the step edges (see inset in Figure 1b) are ascribed to two different effects. The bright line originates from the local phase shift induced by the topographical difference in height. The narrow stripes of light contrast around the step edges are attributed to material contrast which could originate from uncovered SiC areas or already graphene domains. Since graphene growth is rather unlikely at the low growth temperature of 1400 °C [9] an inferior buffer layer growth at the step edges is assumed. The missing buffer layer coverage along the step edges indicate an insufficient carbon supply in these areas which is attributed to carbon diffusion and preferred buffer layer nucleation on the terraces. These line defects separate the buffer layer areas on neighboring terraces, which is unsuitable for the fabrication of large-area QFMLG through intercalation. The missing buffer layer coverage along the step edges indicate an insufficient carbon supply in these areas which is attributed to carbon diffusion and preferred buffer layer nucleation on the terraces.

For higher 100 sccm Ar flow the surface morphology changes. Although the $(6\sqrt{3} \times 6\sqrt{3})R30°$ LEED pattern indicates the formation of the buffer layer on the terraces, Figure 1f, the AFM images of sample $S_{100}$ in Figure 1d, and e show that the smooth terraces are interrupted by canyon-like defects which erode into the SiC terraces and terminate at the following terrace step. These canyon-defects are known to form at gaps in the buffer layer. [21,22] Here the increased Ar flow rate alters locally the thickness and homogeneity of the near surface layer of species (Knudsen layer) [23,24] during the growth, where it causes a faster local SiC decomposition and surface mass diffusion, leading to the canyon-defects before a continuous buffer layer has formed on the terraces.

For much higher Ar flows (Figure.1g and h) the accelerated SiC decomposition induces an etching of the SiC surface. No buffer layer can be formed under these conditions as indicated by the (1x1) LEED pattern of the bare SiC surface, Figure 1i. The AFM image of $S_{1000}$ in Figure 1g shows wide terraces and pronounced terrace broadening and step bunching with step heights of ~2.5 nm. Nanometer-sized islands with triangular shaped basal plane and heights about 5.5 nm are frequently observed on the surface (see supporting information data). A similar trend of Ar flow dependence was observed for typical sublimation growth (SG) without using the PASG technique.[7] (See supporting information)

Our investigations show that with increasing Ar mass flow the SiC surface decomposition is enhanced while the Ar pressure in the reactor is kept constant. This can be understood in a model in which a quasi-thermal equilibrium exists between Si and C species in a surface layer and those in the adjoining gas phase.[23,24] For higher Ar flow the species in the gas phase are increasingly "blown away" by collision processes with the Ar atoms. This perturbation enforces enhanced SiC decomposition to maintain the equilibrium. The decomposition process competes with the buffer layer growth since the C-rich surface reconstruction is known to stabilize the SiC surface by the covalent bonds in-between. [9,22] The final state of the surface is determined by the rates of the involved processes. For zero and small Ar flows the slow SiC decomposition is self-limiting by the generated carbon for buffer layer growth.

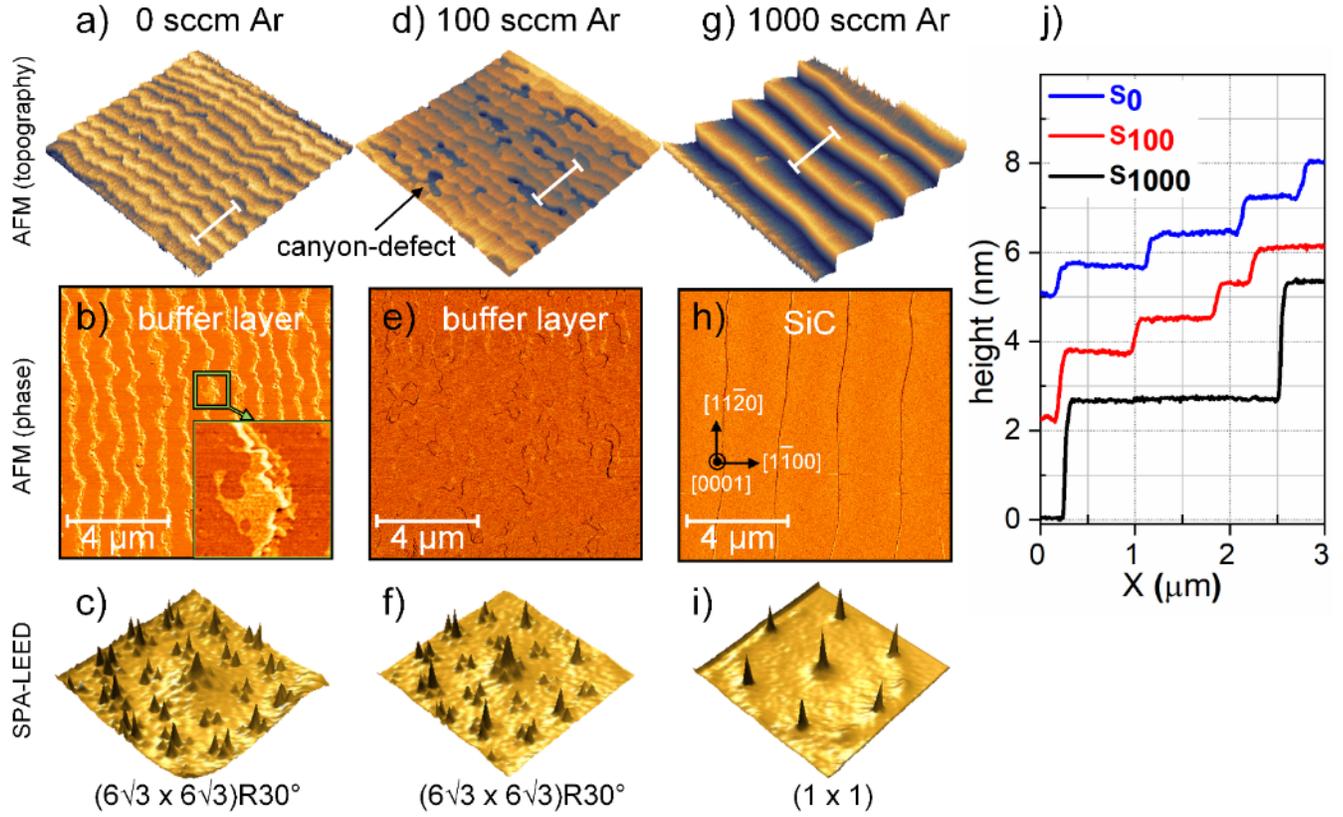

**Figure. 1.** Inspecting the influence of the argon flux on graphitization of 6H-SiC(0001) (1400 °C, 1 bar Ar atmosphere, 30 min) under three different argon mass flows. The AFM topography and phase images are plotted (a), and (b) for $S_0$ (0 sccm Ar), (d) and (e) for $S_{100}$ (100 sccm Ar), and (g), and (h) for $S_{1000}$ (1000 sccm Ar). The inset (1.2 x 1.2 µm²) in (b) shows a close-up of a line defect of sample $S_0$: The lighter narrow stripes are discontinuities in the buffer layer located around the terrace step edge. The step edge itself appears as a very bright line. The dark spots in (d) show canyon defects in the buffer layer on the terraces of sample $S_{100}$. (j) The step-height profiles, extracted from the indicated line in the AFM topography images indicate the giant step bunching under 1000 sccm Ar flow. The LEED images of each sample show typical patterns: A $(6\sqrt{3} \times 6\sqrt{3})R30°$ reconstruction for buffer layer on $S_0$ (c) and $S_{100}$ (f) and a (1 x 1) SiC crystal structure for $S_{1000}$ (i), acquired at 140 eV.

For a high Ar flow, when a fast SiC decomposition rate exceeds the nucleation and growth rate of the buffer layer, an etching of the SiC surface is the consequence. Both extreme cases are displayed by the samples $S_0$ and $S_{1000}$, respectively. For moderate Ar flow both, etching and buffer layer growth can appear simultaneously but spatially separated as seen for $S_{100}$. The control of the SiC decomposition by the Ar mass flow without changing process temperature or the Ar background pressure opens a new parameter range for improved epitaxial graphene growth.

In the following, we demonstrate that the rate of the Ar mass flow also has a substantial impact on the surface morphology of epitaxial monolayer graphene. We compare exemplary two graphene samples, $G_0$, and $G_{20}$, which were grown at 1750°C under 0 and 20 sccm Ar gas flow, respectively. The AFM images for both samples (Figure 2a and b) reveal smooth and regular terraced surfaces covered with monolayer graphene (see Raman spectrum in Figure 4e). Under the slightly increased Ar mass flow of 20 sccm a homogenous step height of 0.75 nm is observed (corresponding to 3 Si-C layers), see cross-section in Figure 2c.

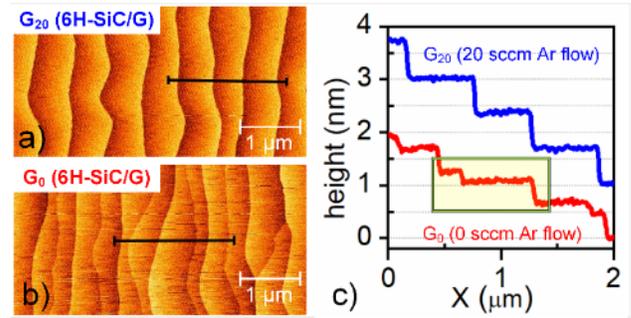

**Figure 2.** Influence of argon mass flow rate on epitaxial growth of graphene on 6H-SiC. (a)(b) AFM topography of two graphene samples, $G_{20}$ and $G_0$, grown under 20 sccm and zero argon flow, respectively, at 1750° and 900 mbar Ar pressure. c) Comparison of AFM height profiles of both samples, $G_0$ and $G_{20}$. The lower step heights and the step pairs of 0.25/0.5 nm (indicated rectangle) exhibit a slower step retraction velocity for growth under zero Ar flow.

For the higher Ar flow a faster decomposition of the SiC layers leads to a step height of 0.75 nm corresponding to half of a 6H-SiC unit cell. The slower SiC decomposition rate of zero Ar flow results in gradually retracting SiC layers. The formation of the observed step pairs of one and two SiC layers are attributed to different retraction velocities of the SiC layers which are related to the inequivalent surface energy of the specific SiC layer sequence of the 6H polytype. [12,13] The retraction process stops when large-area buffer layer coverage on the terraces stabilizes the SiC surface. Once the SiC surface morphology is stabilized by the buffer layer, this structure is frozen and remains stable even when the temperature is further increased for graphene growth to 1750 °C. Accordingly, the SiC morphology is not significantly altered in the subsequent graphene formation process, which is regarded as the formation of the second buffer layer and the detachment and conversion of the first buffer layer into monolayer graphene. The high quality of such ultra-smooth graphene layers was already shown by Raman measurements and nearly isotropic resistivity. [7–9]

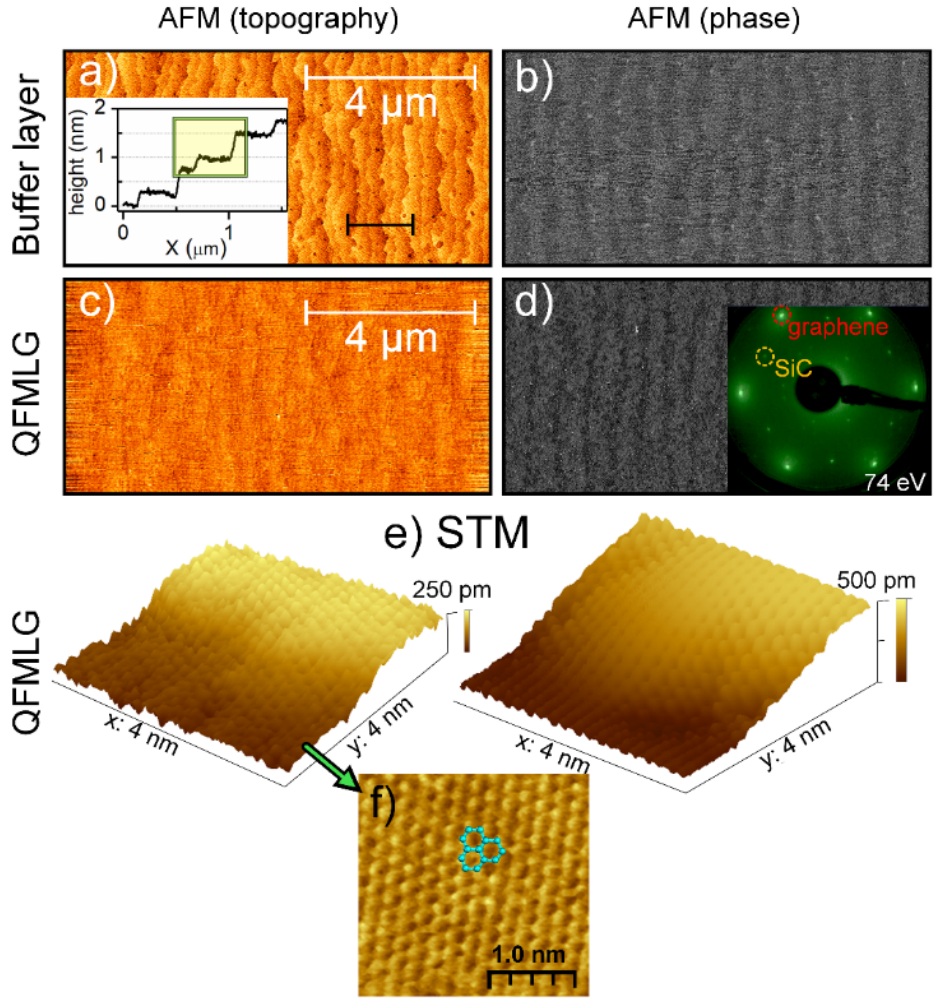

**Figure 3.** AFM measurements of an optimized buffer layer grown on 6H-SiC before and after hydrogen intercalation. (a) shows the ultra-smooth surface of the buffer layer and the pairs of 0.25/0.5 nm steps (inset). (b) The phase image shows a regular contrast pattern on the terraces which is attributed to the different underlying SiC layer sequence. (c) After H intercalation a coherent sheet of QFMLG is obtained. (d) The regular contrast pattern is still visible in the phase image. The inset shows the typical LEED pattern of quasi-freestanding graphene. (e) Scanning tunneling microscopy (STM) inspection (4 x 4 nm$^2$) of the QFMLG near terrace steps of minimum feasible step-height of 0.25 and 0.5 nm show perfect coverage with the single freestanding graphene layer. (f) The atomic resolution topography of the QFMLG in the marked square terrace area is obtained by constant-current STM (1 nA, 0.1 V). Three hexagonal carbon rings (cyan) are indicated.

**Quasi-freestanding mono/bi- graphene layers.** In the following, we focus on the investigation of the coherent high-quality buffer layer which can be used for the fabrication of large areas of QFMLG. Moreover, the investigation of the quasi-freestanding graphene layers is a further test for the homogeneity of the produced buffer layer. To this end, we started with the growth condition of $S_0$ but with an optimized time and annealing protocol as given above. The AFM topography of this optimized buffer layer sample $B_0$ in Figure 3a shows very smooth buffer layer with step heights below 0.75 nm (inset in Figure 3a) and a repeating pattern of step pairs of 0.25 and 0.5 nm which indicates a reduced step retraction compared to the samples $S_0$ and $S_{100}$. No canyon defects appear in this sample and the corresponding phase image in Figure 3b also shows no line defects which indicates a continuous buffer layer which spans over the terrace edges. The small step heights are supposed to be additionally beneficial for the linking process of the buffer layer on neighboring terraces. Although the surface is completely covered with the buffer layer, a weak contrast between both alternating terraces is observed which is attributed to the influence of the underlying SiC layers. (to be published)

The structural homogeneity and lateral coverage of the buffer layer $B_0$ are further verified by Raman measurements. The spectrum of this sample shows broad features, between 1200 and 1700 cm$^{-1}$ (upper spectrum in Figure 4a) which are related to the vibrational density of states (vDOS) of the $(6\sqrt{3} \times 6\sqrt{3})R30°$ surface reconstruction.[25] The integrated intensity area of these broad Raman bands is regarded as a measure for lateral coverage in the Raman mapping of the buffer layer, and it is plotted in Figure 4b for an area scan of 20 x 20 µm$^2$. Additionally, no Raman spectral changes were observed in the vDOS bands during the Raman mapping indicating a homogenous distribution of the buffer layer across the investigated area. The nearly monochrome green colored area visualizes that a continuous and homogenous buffer layer has formed. This becomes obvious when the Raman map is compared to that of a graphene sample which was grown under non-optimized conditions, see supporting information. There, the spatial variation of the integrated buffer layer intensity displays a considerable non-homogenous coverage and the partial lack of the buffer layer. The spectra of the Raman mapping show no graphene-typical 2D peak (at around 2700 cm$^{-1}$) which clearly indicates the absence of EG domains on top of the buffer layer.

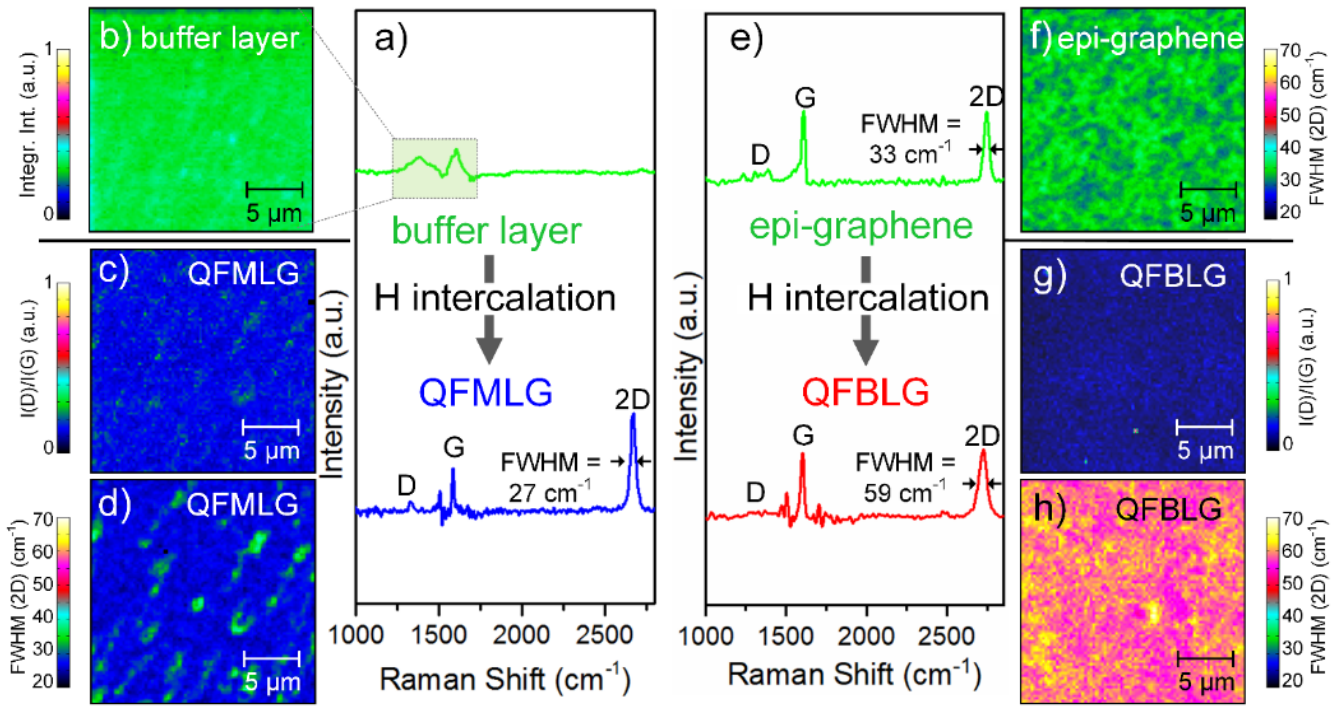

**Figure 4.** Micro-Raman spectroscopy of graphene samples before and after hydrogen intercalation. (a) shows the Raman spectra of an optimized buffer layer sample (upper spectrum) and the resulting QFMLG (lower spectrum) after hydrogen intercalation. In (b) the 25 µm x 25 µm map the integrated intensity of the buffer layer Raman band is plotted. In (c) the intensity ratio of D- and G- peak (peak values) and in (d) the linewidths (FWHM) of the 2D peak of the QFMLG sample are displayed in areal maps. (e) shows Raman spectra of epitaxial monolayer graphene and the resulting QFBLG obtained after hydrogen intercalation. (f) The linewidths (FWHM) of the 2D line of the epitaxial graphene layer shows homogenous monolayer graphene without bilayer inclusions. (g) Areal maps of the intensity ratio of D- and G- peak (peak values) and (h) linewidths (FWHM) of the 2D peak of the QFBLG sample.

QFMLG was produced by hydrogen intercalation of the optimized buffer layer ($B_0$) and the AFM, LEED, and STM scrutiny of the surface are shown in Figure 3c-f. The AFM image in Figure 3c reveals a homogenous QFMLG layer. The detachment of the graphene leads to a smoother surface compared to the previous buffer layer which showed sharper and more pronounced steps in the AFM image (Figure 3a). The large area detachment of the buffer layer is proven by the typical LEED pattern (inset in Figure 3d) of quasi-freestanding graphene.[1] The $(6\sqrt{3} \times 6\sqrt{3})R30°$ buffer layer pattern has disappeared since the correlation of the buffer layer superstructure to the underlying SiC surface lattice is lost. Interestingly, the AFM phase image (Figure 3d) is still showing a slight alternating contrast on the terraces which reveals a continuous influence of the underlying SiC surface terraces on the electronic properties of the QFMLG.

Additionally, the atomic structure was investigated using an Omicron low-temperature STM at 77 K with a tungsten tip. The detailed topography of the QFMLG sheet near 0.25 nm high and 0.5 nm high step edges are displayed by high-resolution STM images in Figure 3e. The graphene lattice is clearly seen on both the upper and lower terraces. The observation of the hexagonal crystal structure (lattice constants of 2.46 Å) proves that the SiC in this area is completely covered with graphene and line defects are absent. Furthermore, dislocations and domain boundaries are not observed. The 4 x 4 nm$^2$ STM images which were taken across the step edges reveal a coherent graphene layer which spans smoothly over the step edge from one terrace to the other. A larger two-dimensional image (see supporting information) shows the unchanged lattice orientation over the step. Our finding is similar to the case of monolayer graphene covered step. [26,27] This is also very similar to the case of QFMLG on the higher step of 0.75 nm which randomly was observed on the surface of this sample, see also supporting information. However, in our QFMLG the warp up (down) of the graphene sheet at the upper (lower) terrace as observed for EG [26,27] is not formed.

The appearance of the 2D peak in the Raman spectra of the QFMLG sample (lower spectrum in Figure 4a) proves that quasi-freestanding monolayer graphene (QFMLG) has been produced, in agreement with LEED shown in the inset of Figure 3d. The relaxation of the graphene layer is indicated by the redshift of the 2D peak position at 2669 ± 2.7 cm$^{-1}$ compared to the 2D peak position at ~2731 ± 1.5 cm$^{-1}$ of monolayer epitaxial SiC/G van-der-Waals bonded to the buffer layer. From the small full-width-at-half-maximum (FWHM) value of 27 ± 2.2 cm$^{-1}$, a high carrier mobility value is expected. [28]

The Raman spectrum of the QFMLG sample in Figure 4a (lower spectrum) shows a well pronounced G peak (1587 ± 1.1 cm$^{-1}$) with an FWHM of 9.6 ± 1.9 cm$^{-1}$. A very small D peak at ~1339 cm$^{-1}$ in the Raman spectrum of QFMLG can be attributed to a small density of remaining defects in the graphene lattice. The small $I_D/I_G$ (peak maxima) ratio of about 0.1 is comparable to that of other high-quality graphene samples. [29] From the $I_D/I_G$ ratio a defect density of $n_{def.} = (3.3 ± 0.7) \times 10^{10}$ cm$^{-2}$ is estimated. [29,30] This extraordinarily high graphene quality was found over the entire area of 20 x 20 μm$^{-2}$ in the mapping of the $I_D/I_G$ (peak values) ratio in Figure 4c. Measurements at different positions and on other optimized buffer layer samples suggest a similar low defect density over the whole sample surface. The origin of the defects is correlated with the SiC crystal imperfections that induce defects in the buffer layer during growth, as well as the intercalation deficiency like partial intercalation or etching which could have reasonably been enhanced by low concentration of the applied hydrogen. [26,27,31]

In Figure 4a and e, the Raman spectra were subtracted from a reference Raman spectrum of pure SiC to remove the spectral overtones related to SiC. A spectral artifact (wiggles) appear (in blue and red spectra) at ~1500 cm$^{-1}$ and ~1700 cm$^{-1}$ which is due to a slight mismatch between the Raman spectrum of the samples and the reference spectrum of SiC.

The Raman mapping of the FWHM of the 2D peak in Figure 4d shows predominantly blue marked regions that are related to 2D peak widths of ~27 cm$^{-1}$ which give evidence of a very high homogeneity QFMLG. Bilayer formation, in this case, is excluded since it would result in much larger FWHM values > 45 cm$^{-1}$. [28] The small green colored areas show increased FWHM values slightly above 30 cm$^{-1}$ which could arise from low strain variations at nanoscale leading to a superposition of slightly different 2D peak positions within the Raman laser spot and thus exhibiting an artificial broadening of the 2D peak width in the acquired Raman spectrum and mapping.

Also, QFBLG was produced by hydrogen intercalation from the optimized monolayer graphene sample ($G_0$). The upper Raman spectrum in Figure 4e shows the typical fingerprint of epitaxial graphene, indicating the G peak at 1601 cm$^{-1}$ and 2D peak at 2731 cm$^{-1}$, whereas broad phonon bands from the buffer layer arise in the range of 1200 and 1700 cm$^{-1}$. Figure 4f shows a homogenous distribution of the 2D peak width of epitaxial graphene over an area of 20 μm x 20 μm with an averaged 2D peak width of (33 ± 1.5 cm$^{-1}$). After H intercalation (lower spectrum in Figure 4e) the broad buffer layer related Raman band around the D peak disappears since the detached buffer layer is transformed into the second free-standing graphene layer. Therefore, the 2D peak becomes broader (FWHM around 59 cm$^{-1}$) and asymmetric line shape (see supporting information) which indicates the formation of bilayer graphene. The FWHM map of the 2D peak in Figure 4h reveals the homogenous distribution of bilayer graphene. The slightly increased FWHM values (yellow areas) could again be caused by local strain variations. The quality and homogeneity of the QFBLG is further underlined by the low values and even distribution of $I_D/I_G < 0.1$ (peak maxima) ratios which indicate a low defect density of about $n_{def.} < 2.0 \times 10^{10}$ cm$^{-2}$, see Figure 4g.

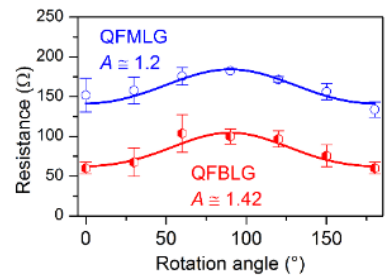

**Figure 5.** Resistance measurements by rotational four-point probe measurements (100 μm x 100 μm) as a function of rotation angle for QFMLG and QFBLG samples. A rotation angle of 0° corresponds to transport parallel to the terraces and 90° rectangular to the step edges. The anisotropy values $A$ are given as calculated from fit curves of $\rho_{perp} / \rho_{par}$.

**Electronic transport characterization.** The homogeneity of the QFMLG and the QFBLG samples were further investigated by rotational four-point probe measurements on the micro scale (100 μm STM tip spacing) at room temperature as described in Ref. [8,32–34]. The angle-dependent resistance of both samples is plotted in Figure 5. The lowest resistance values for each sample are measured for transport parallel to the terraces whereas the maximum value is obtained at an angle of about 90° which corresponds to a current direction perpendicular to the step edges. [8,34] The calculated anisotropy values ($A = \rho_{perp} / \rho_{par}$) are 1.2 and 1.42 for QFMLG and QFBLG, respectively. Despite missing comparative values in the literature these values are regarded as a sufficient low anisotropy indicating a good homogeneity. However, they are larger compared to optimized ultra-smooth epitaxial monolayer graphene with nearly unity isotropy values ($A = 1.03$). [8] We suppose that the observed anisotropy could stem from intercalation related defects and local strain variation as for instance observed in the Raman spectrum of the QFMLG sample, see Figure 4c. The low concentration of the intercalating gas agent (5% hydrogen) could have also in tensified the defects and result in an inhomogeneous or partial intercalation. [35,36]

The microscopic anisotropy measurements were complemented by macroscopic VdP measurements in a helium flow cryostat in a magnetic field up to 500 mT. For the measurement the samples were first cut into square-shape of 5 mm x 5 mm, and the graphene on top of the sample was isolated from the graphene on the side and the back of the substrate by scribing cut-grooves on each side close to the edge of the sample (~ 0.1 mm from the edge), as shown in the schematic of the VdP configuration in Figure 6.

| sample | T (K) | $R_{sheet}$ (Ωsq) | $R_{par}$ (Ω) | $R_{perp}$ (Ω) | $A = \frac{R_{perp}}{R_{par}}$ | p, n ($10^{12}$ cm$^{-2}$) | μ (cm²/Vs) |
|---|---|---|---|---|---|---|---|
| QFMLG (B₀) | 295 | 812 | 164 | 194 | 1.18 | p = 6.7 | 1159 |
| | 2.2 | 837 | 174 | 195 | 1.16 | p = 6.4 | 1169 |
| Epi-graphene (G₀) | 295 | 979 | 209 | 219 | 1.05 | n = 6.8 | 1108 |
| | 2.2 | 365 | 79 | 82 | 1.04 | n = 6.9 | 2459 |
| QFBLG (G₀) | 295 | 364 | 68 | 93 | 1.37 | p = 6.3 | 2696 |
| | 2.2 | 345 | 64 | 90 | 1.4 | p = 5.4 | 3352 |

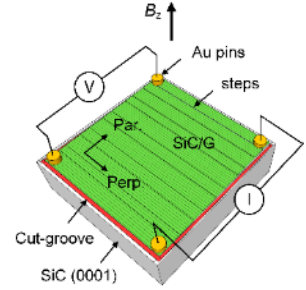

**Table 1.** Results of VdP measurements at 5 mm x 5 mm large samples with QFMLG, epitaxial graphene, and QFBLG.

**Figure 6.** Right side: sketch of the VdP configuration. The graphene (green) on top of the sample (5 x 5 mm2) was isolated from graphene on sidewalls and backside by cut-grooves close to each edge (indicated as red lines).

Four gold pins in square configuration were pressed firmly onto the surface to contact the graphene at the corners. The ohmic characteristic and the linearity of the Hall ramps were tested before the measurements. The VdP measurements were carried out at room temperature (295 K) and 2.2 K., see Table 1. As expected both samples, QFMLG and QFBLG, show the typical high p-type carrier density of about 6.7 x $10^{12}$ and 6.3 x $10^{12}$ cm$^{-2}$, respectively. [1] For the QFMLG we obtain mobility of about 1159 cm²/Vs at room temperature and for n-type epitaxial graphene (G₀) about 1108 cm²/Vs. However, at 2.2 K the epitaxial graphene shows considerably higher mobility of 2459 cm²/Vs whereas the mobility of QFMLG remained almost constant at $\mu = 1169$ cm²/Vs. This is due to different scattering mechanisms in these two types of epitaxial graphene monolayers. While the temperature dependence of mobility in the epitaxial graphene (G₀) is due to longitudinal acoustic phonons in graphene and SiC, the dominant scattering in QFMLG is attributed to Coulomb scattering induced by charged impurities. [6,37] Although higher values of 3000 cm²/Vs were already reported for micrometer-sized QFMLG confined Hall bars on the single terrace [2,6,14] this is still a remarkable result, for the produced large-size sample. In contrast to QFMLG, the measurements on QFBLG (after intercalating G₀) revealed noticeable higher mobility values of 2696 cm²/Vs (RT) and 3352 cm²/Vs (2.2 K). This temperature dependence of the carrier mobility in QFBLG is referred to an interplay of different scattering mechanisms, temperature dependent Coulomb scattering and the charge impurity density. Note, that the higher mobility compared to QFMLG could stem from screening of Coulomb scatterers in the substrate and bilayer graphene. [6,38,39]

It is worth mentioning that investigation of QFMLG and QFBLG was carried out on a considerable number of sample sets showing much higher mobilities of 1300 cm²/Vs, 2000 cm²/Vs, and 3300 cm²/Vs for QFMLG, epi-graphene, and QFBLG, at room temperature, respectively. However, for consistency, we present in this work the results of the samples with similar treatment and the same growth and intercalation processes.

The homogeneity of the samples was derived from the VdP sheet resistances $R_{par}$ and $R_{perp}$ measured in two orthogonal directions, parallel and perpendicular to the step edges. As before higher resistance values were obtained for perpendicular transport. The QFMLG and QFBLG samples show anisotropy values ($R_{perp} / R_{par}$) values of 1.18 and 1.37 (at room temperature) which are in very good agreement with the microscopic four-point probe measurements. This result indicates the very good electronic homogeneity over mm scales of our QFMLG and QFBLG samples in excellent agreement with the AFM and Raman data. For QFBLG a comparable VdP study [19] revealed a much stronger anisotropy of about 200% which was attributed to high step edges (~ 10 nm) and multilayer graphene along the step edges. The absence of multilayer graphene in our QFMLG and QFBLG and the low step heights can thus be regarded as highly beneficial for homogeneous electronic properties. This result gives evidence of the high quality of the PASG method and the possibility of optimization of the growth parameters.

## CONCLUSION

In summary, we have presented AFM, STM, LEED, electronic transport and Raman measurements which indicate the strong influence of the argon mass flow rate on the formation of the buffer layer and the graphene growth. For a given temperature and constant Ar pressure, the Ar mass flow rate controls the SiC decomposition rate which can be qualitatively understood by thermal equilibrium considerations. This new finding has the potential to improve the graphene quality by avoiding accelerated step bunching at higher temperatures and graphene roughening for lower Ar pressures, respectively. By properly chosen growth parameters it is possible to prevent structural defects (canyon defects and step defects) and to obtain a continuous, large-area buffer layer without graphene inclusions. This meets the demands of intercalation purposes and could further be used as a platform for growing other 2D materials or metamaterials. The QFMLG and QFBLG produced by hydrogen intercalation exhibit excellent homogeneity and very small resistance anisotropy over areas in the millimeter range. This indicates the presence of a coherent layer of free-standing monolayer graphene over large areas which is a prerequisite for application in electronic devices.

## ASSOCIATED CONTENT
**Supporting Information.** The Supporting Information is available at the end of this document.

Corresponding Author
*E-mail: davood.momeni.pakdehi@ptb.de
*E-mail: klaus.pierz@ptb.de

**Acknowledgment.** D.M P. acknowledges support from the School for Contacts in Nanosystems (NTH nano). J.A. thanks the support from DFG project Te386/12-1.

# Supporting Information

## Homogeneous Large-area Quasi-freestanding Monolayer and Bilayer Graphene on SiC

D. Momeni Pakdehi,[*,1] K. Pierz,[*,1] S. Wundrack,[1] J. Aprojanz,[2] T. T. N. Nguyen,[3] T. Dziomba,[1] F. Hohls,[1] A. Bakin,[4,5] R. Stosch,[1] C. Tegenkamp,[2,3], F. J. Ahlers,[1] and H. W. Schumacher[1]

[1]Physikalisch-Technische Bundesanstalt, Bundesallee 100, 38116 Braunschweig, Germany
[2]Institut für Festkörperphysik, Leibniz Universität Hannover, Appelstraße 2, 30167 Hannover, Germany
[3]Institut für Physik, Technische Universität Chemnitz, Reichenhainer Straße 70, 09126 Chemnitz, Germany
[4]Institut für Halbleitertechnik, Technische Universität Braunschweig, Hans-Sommer Straße 66, 38106 Braunschweig, Germany
[5]Laboratory of Emerging Nanometrology (LENA), Technische Universität Braunschweig, Langer Kamp 6a, 38106 Braunschweig, Germany

*E-mail: davood.momeni.pakdehi@ptb.de
*E-mail: klaus.pierz@ptb.de

**1) Influence of argon flux on conventional epitaxial sublimation growth (SG) of buffer layer**

Figure S1 shows the influence of the argon mass flow rate on buffer layer growth on the samples without polymer preparation that are investigated by AFM and SEM. Three samples S'$_0$, S'$_{100}$, and S'$_{1000}$ are 4H-SiC with nominal miscut of about -0.06° towards [1$\bar{1}$00] and were processed at 1400 °C (1 bar argon atm., for 30 min) at different Ar flow of 0, 100 and 1000 sccm, respectively. This experiment was carried out under the same conditions as the one in the manuscript and it aims at studying the influence of argon flow on samples grown from another SiC polytype (4H-SiC) in the absence of polymer preparation. For the high argon flow of 1000 sccm the surface of the sample undergoes severe step bunching without any buffer layer growth Figure S1. g, h, i. This is similar to the PASG sample in the manuscript (see Figure 1g, h, and i). In both cases the high Ar flow leads to surface etching. For moderate Ar flow, however, the situation is different. The surface of the sample without polymer preparation shows stripes of the covered buffer layer and bare SiC, Figure S1. d, e, and f. This is in contrast to the PASG sample (manuscript Figure 1, d, e, and f) where the provided carbon species from polymer lead to surface super-saturation and well buffer layer coverage although the Ar flow caused canyon-like defects. For the case of zero argon flow (Figure S1, a, b, and c) the surface looks very good with homogeneous coverage, while the terraces appear less ordered in comparison with the PASG sample (manuscript Figure 1, a, b, and c).

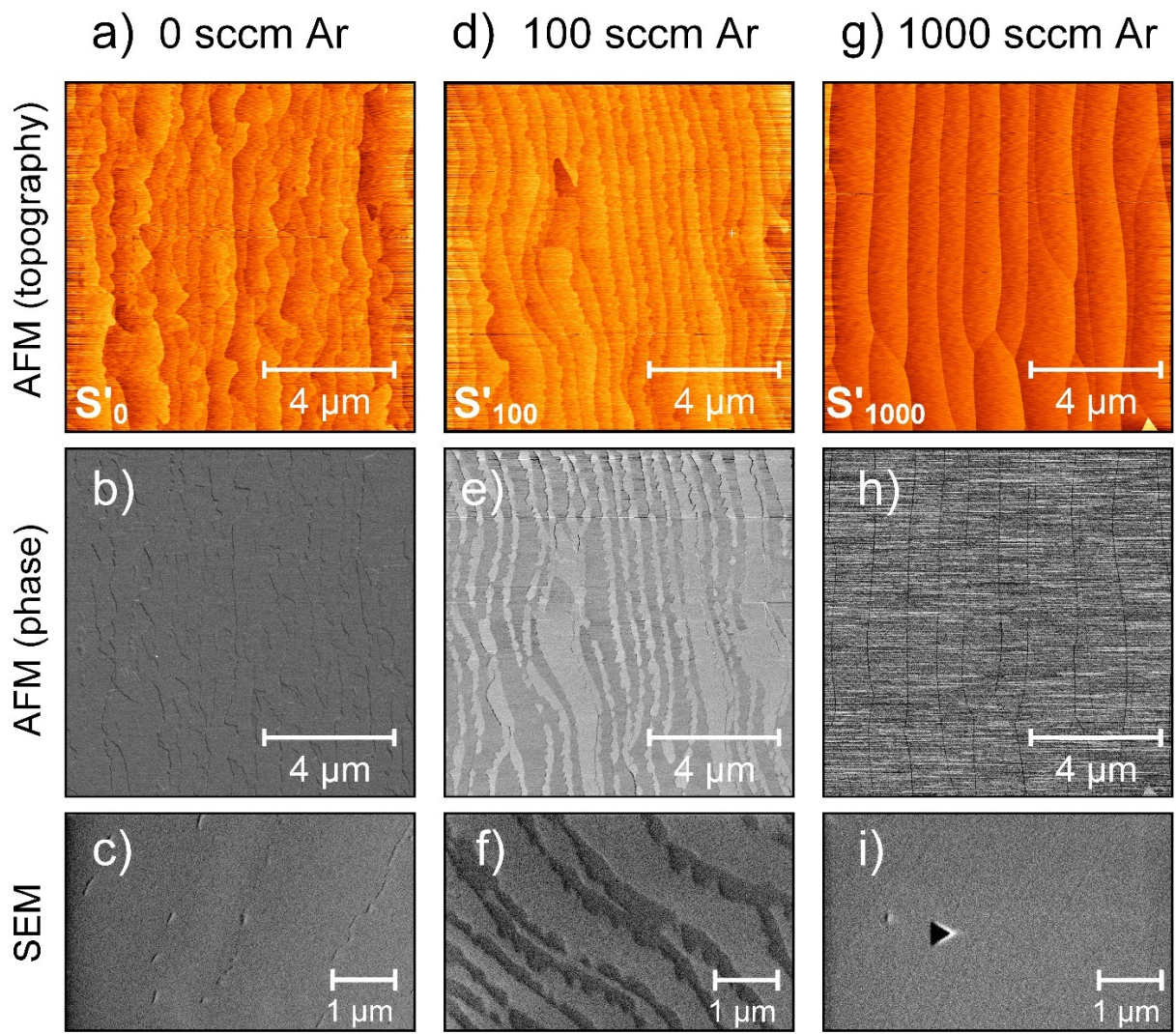

**Figure S1.** Inspecting the influence of the argon mass flow on graphitization of 4H-SiC(0001) at 1400°C (1bar Ar ambient, 30 min) under three different argon gas flows: a) S'$_0$ (Ar/ 0 sccm), b) S'$_{100}$ (Ar/100 sccm), and c) S'$_{1000}$ (Ar/1000 sccm). The sample was grown by typical sublimation growth without polymer preparation. S'$_0$ processed under no Ar flow representing good buffer layer coverage in AFM phase (b) and scanning electron microscopy SEM (1kV) (c) images. The moderate flow Ar for S'$_{100}$ distorts its surface growth causing the formation of buffer layer stripes on this sample, as can be seen in AFM phase (e) and SEM (f) images. The intensive argon flux on S'$_{1000}$ prevents buffer layer formation and results in severe step bunching on this sample (g), (h), (i).

### 2) Formation of triangle-shape defects

It is observed that the increase of the Ar flow leads to the formation of triangular-shape structures. This can be seen in Figure S1g, h, and i, for the sample processed under 1000 sccm Ar flow. Also, rather increase of the Ar flow escalates the density of such structures, as can be seen in Figure S2 for the sample processed at 1400 °C (30 min, 1 bar Ar) in the presence of the Ar flow of 2000 sccm. The aggregated mass along the giant steps and the triangular-like structures is the typical morphology all over the surface of this sample. Although, here the properties of such triangular-shape structures has not been further studied, however, they very resemble the cubic SiC grown on other substrates elsewhere. [1,2]

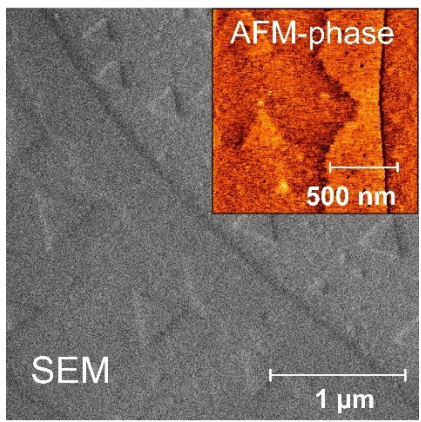

**Figure S2.** Formation of triangular-like structures at high argon gas flow. Scanning electron micrograph (1kV) of a 6H-SiC sample after annealing at 1400°C (1 bar in Ar ambient, 30 min) under 2000 sccm Ar-flux. The inset shows the AFM phase image of the same sample.

Moreover, such triangle–shape structures might not only appear under intensive gas flow but also under the lower pressure conditions in, which leads to condensation and reflection of sublimated species back onto the surface of the substrate. Figure S3 shows AFM inspection on the surface of a 6H-SiC sample which processed at 50 mbar (1400 °C, 30 min). The triangle-bar structures appear to entirely cover the substrate with heights up to 14 nm.

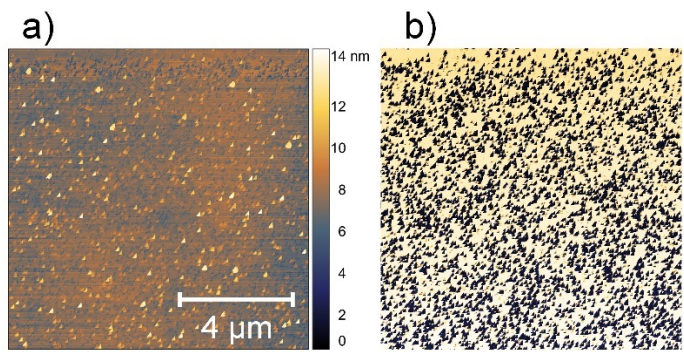

**Figure S3.** AFM inspection: (a) morphology (b) phase image. Triangular-shape structures appeared on the surface of a sample processed at 1400 °C (30 min), but the pressure was 50 mbar instead of the typical 1 bar of argon. The surface shows high condensation of the sublimated species back on the surface of the sample.

### 3) Step-by-step transition of the buffer layer to quasi-freestanding monolayer graphene

Figure S4a shows the Raman spectra of a buffer layer sample after step-by-step hydrogen intercalation 5% (95% argon) for 15 minutes at different temperatures ranging from 400 °C up to 1200 °C. The intercalation up to 400 °C does not show any significant change in the buffer layer spectra with the typical two-phonon bands in the spectral range between 1390 and 1605 cm$^{-1}$. [3] By temperature risse to 500°C despite the slight change in D-peak, yet no 2D-peak is observed, whereas at 600°C the D-peak sharply increases and the 2D-peak is observed, clearly indicating a transition from buffer layer to graphene. By further increasing temperature, the D-peak decreases, which denotes disorder reduction in the crystal of graphene. An even higher temperate leads to a higher 2D-peak, while its position shifts to lower wave numbers. This is a result of a tensile strain which may be accompanied with a doping increase.[4,5] Moreover, the increase of the temperature shows lowering down the defect density of the QFMLG layer as shown for three annealing steps of 600 °C, 800 °C, and 1000 °C, in Figure S4b, c, and d.

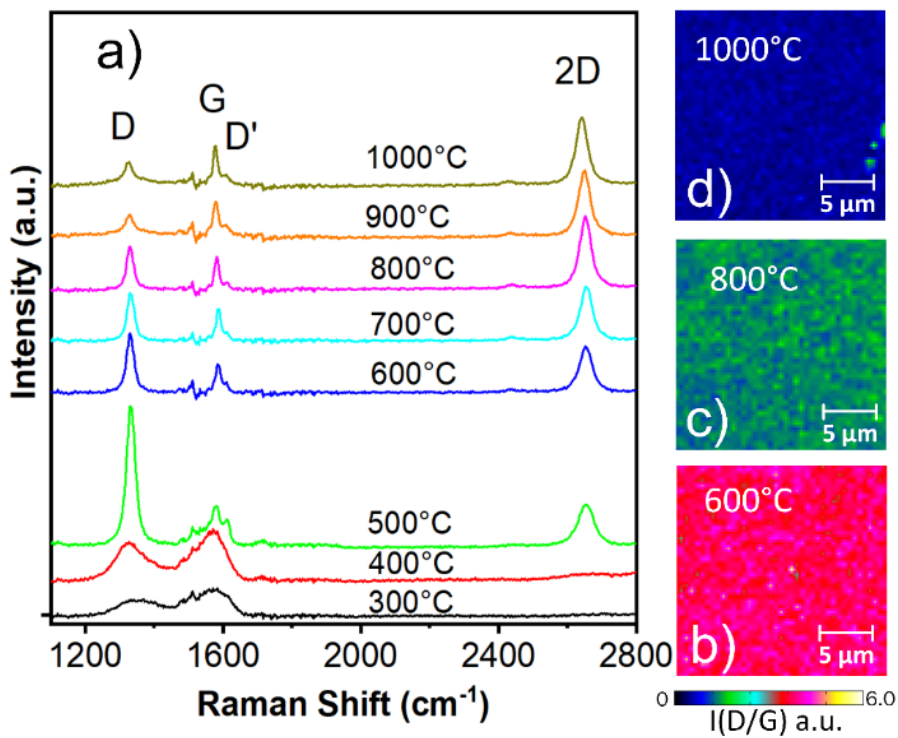

**Figure S4.** Temperature dependency of intercalation on buffer layer sample via 5% (95%Ar) hydrogen. At 400°C no intercalation occurred, inferred from the Raman spectra (a). For temperature above 400°C, the D-peak starts to increase which is accompanied by the appearance of 2D-peak at 600°C, indicating transition to graphene. By increasing the temperature, D-peak decreases and 2D-peak increases demonstrating more effective intercalation and reduction of defect density measured by $I_D/I_G$ (b-d).

**4) Hydrogen intercalation on low-quality buffer layer**

Figure S5 shows the AFM and Raman investigations of a buffer layer sample with poor coverage. The AFM phase images (Figure S5b) a color contrast on the abutting terrace, though the origin of the contrast is different from the optimized sample in the manuscript (see Figure 3b). The contrast on this sample is a result of interaction between AFM tip with two different materials (buffer layer and SiC stripes) directly at the sample surface, whereas the contrast on the optimized sample is attributed to the underlying SiC surfaces. The mapping Raman D+G areas in Figure S5d shows an inhomogeneous buffer layer. As it was expected, the hydrogen intercalation on such defected buffer layer has led to a highly defected QFMLG as shown in the Raman mapping of $I_D/I_G$ in Figure S5f.

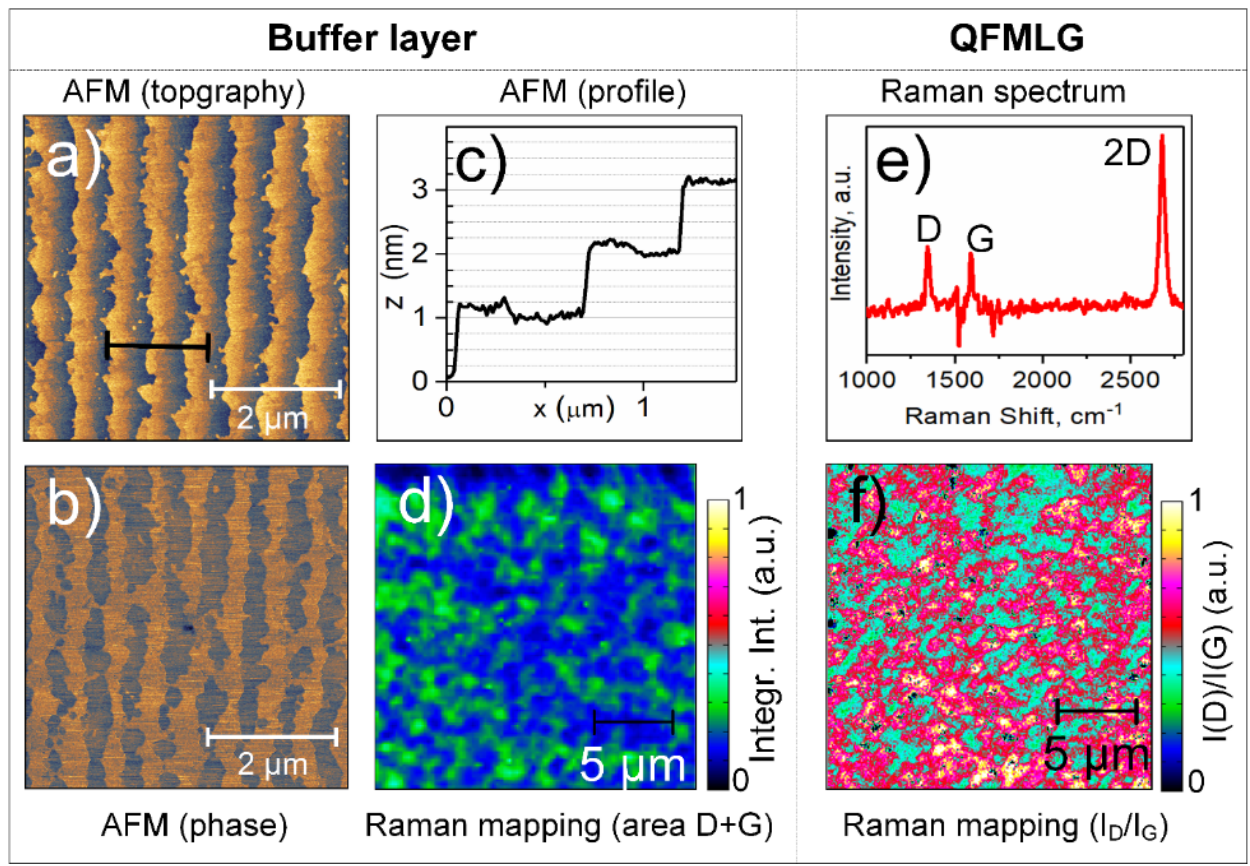

**Figure S5.** AFM topography (a), height profile (b) and phase (c) images of a buffer layer sample with poor coverage showing a phase color contrast corresponding to SiC and buffer layer stripes. (d) Raman mapping of D+G peaks area of the buffer layer. (e) Raman spectrum of the sample after the intercalation by hydrogen 5% (95% Ar). f) Raman Mapping of $I(D/G)$ shows an inhomogeneous QFMLG.

### 5) Atomic resolution STM measurement on QFMLG

Figure S6a shows a two-dimensional STM image captured on QFMLG sample shown in the manuscript (see Figure 3e) with a larger size of 15 x 15 nm². It can be seen that the quasi-freestanding monolayer graphene very smoothly passes over the step of 0.25 nm high from one terrace to the neighboring terrace. Figure S6b exhibits the QFMLG nicely covers a higher step of 0.75 nm high (1/2 unit-cell of 6H-SiC) which randomly was observed on this sample.

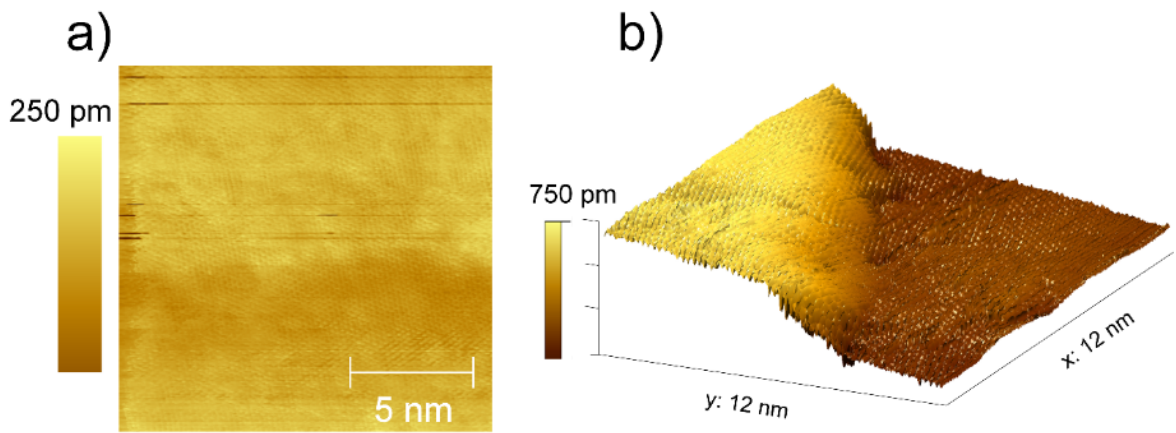

**Figure S6.** STM topography inspection of QFMLG (shown in the manuscript) on (a) a terrace with step of 0.25 nm high and (b) a terrace with step height of 0.75 nm.

**6) 2D-peak spectrum of Quasi-freestanding bilayer graphene (QFBLG)**

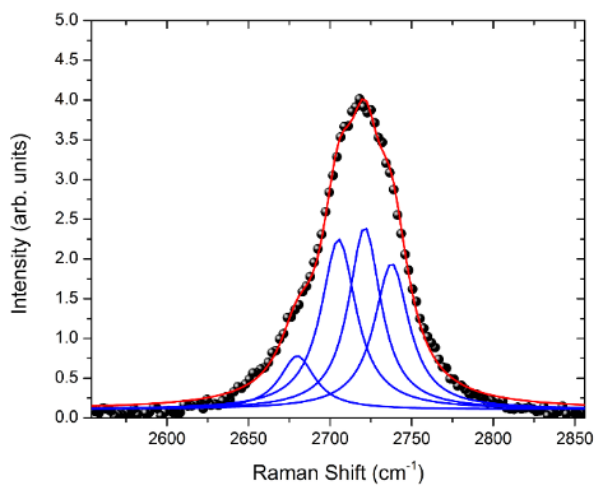

**Figure S7.** Raman spectrum of QFBLG shows broadened 2D peak with an FWHM of ~ 59 cm$^{-1}$ as well as an asymmetrical line shape of the 2D peak which can be fitted by four Lorentzian curves.